\documentclass[prl,showpacs,twocolumn,letter,
               groupedaddress,floatfix,10point]
              {revtex4}
\usepackage[utf8]{inputenc}
\usepackage{epsfig}
\usepackage[T1]{fontenc}
\usepackage{graphicx}
\usepackage{amssymb}
\usepackage{amsmath}

\usepackage{relsize}
\usepackage{color}
\usepackage{times}
\usepackage{bm}
\usepackage[section]{placeins}
\usepackage{array}
\newcommand{\floor}[1]{{\lfloor #1\rfloor}}
\newcommand{\ceil}[1]{{\lceil #1\rceil}}

\begin{document}

\title{Redundancy and error resilience in Boolean Networks}

\author{Tiago P. Peixoto}
\email[]{tiago@fkp.tu-darmstadt.de}
\affiliation{Institut für Festkörperphysik, TU Darmstadt, Hochschulstrasse 6,
  64289 Darmstadt, Germany}

\date{\today}

\begin{abstract}
  We consider the effect of noise in sparse Boolean Networks with redundant
  functions. We show that they always exhibit a non-zero error level, and the
  dynamics undergoes a phase transition from non-ergodicity to ergodicity, as a
  function of noise, after which the system is no longer capable of preserving a
  memory of its initial state.  We obtain upper-bounds on the critical value of
  noise for networks of different sparsity.
\end{abstract}

%\keywords{}

\maketitle

\emph{Introduction}--- Biological systems are unavoidably noisy in their nature,
but often need to function in a predictable
fashion~\cite{metzler_keeping_2009}. In such a situation, strategies to diminish
the harmful effect of noise will significantly impact the fitness of a given
organism. The most fundamental protection mechanism a system can adopt is the
redundancy of its underlying components, since the resulting coincidences
necessary to impact the proper function of the system can drastically diminish
the probability of error. In this letter we are concerned with the effect of
redundancy in gene regulation; in particular in a simple Boolean Network (BN)
model. We assume that each component in the system is arbitrarily redundant,
with the only restriction that the number of inputs per component is fixed and
finite. In a general manner, we are able to show that redundancy can always
guarantee reliable dynamics, up to a given critical value of noise, above which
the system is incapable of maintaining any memory of its past states. From
simple considerations, we are able to obtain upper bounds on the maximum
resilience attainable. This provides an important frame of reference to
determine the reliability of a system with a given sparsity.

We begin by defining the model, and how noise is introduced. A Boolean Network
(BN)~\cite{kauffman_metabolic_1969} is a directed graph with $N$ nodes,
representing the genes, which have an associated Boolean state $\sigma_i \in
\{0,1\}$, corresponding to the transcription state, and a function
$f_i(\{\sigma_j\}_i)$, which determines the state of node $i$ given the states
of its input nodes $\{\sigma_j\}_i$. The number of inputs of a given node is
$k_i$, or simply $k$ if its the same for all nodes. This system is usually
updated in parallel, such that at each time step $t$, we have $\sigma_i(t+1) =
f_i(\{\sigma_j(t)\}_i)$. Starting from an initial configuration, the system will
evolve, and eventually settle on an \emph{attractor}. In a real system, the
expression level of a particular gene can fluctuate, despite the stability of
its input states~\cite{mcadams_stochastic_1997}. This characteristic can be
incorporated qualitatively in the BN model as uniform
noise~\cite{miranda_noise_1989,
golinelli_barrier_1989,aleksiejuk_ferromagnetic_2002, huepe_dynamical_2002,
qu_numerical_2002, indekeu_special_2004, peixoto_noise_2009}, defined as a
probability $p$ that, at each time step, the value of a given input $\sigma_j
\in \{\sigma_j\}_i$ of a node $i$ is flipped, prior to the evaluation of the
function $f_i$. The value of $p$ plays the role of a temperature in the
system. If $p=0$ the original deterministic model is recovered, and if $p=1/2$
the system becomes effectively decoupled, with entirely stochastic dynamics.

In the model above, it is known that error resilience does not spontaneously
emerge, since Random Boolean Networks (BNs with random topology and
functions~\cite{drossel_random_2008}), and simple functional elements such as
loops always exhibit ergodic behaviour in the presence of noise
($p>0$)~\cite{peixoto_noise_2009}. To obtain resilience, some level of
functional redundancy must be introduced in the network. In the following we
describe how this can be done, and analyse the optimal situation where all
functions are arbitrarily redundant. From this situation we obtain upper bounds
on the maximum reliability attainable, which is characterized by a transition
from non-ergodic to ergodic behaviour at a critical noise value.

\emph{Redundancy in sparse networks}--- Given a finite number of inputs per node
$k\ll N$, for any given (non-constant) function, the probability that noise will
change the output of the function will always be above zero, independent of the
size of the network. Therefore, in average there will always be a non-vanishing
fraction of the nodes which will be at the wrong state at any given time. The
most that can be expected is that this fraction be as small as possible, and
\emph{remain} small as the dynamics evolve. The issue of achieving the first
goal was first approached by von Neumann~\cite{von_neumann_probabilistic_1956,
pippenger_developments_1990}, who described a general mechanism of optimal
redundancy, which is capable of reducing the propagation of errors in a BN. We
will briefly outline this mechanism, and then show how it can be used to
construct a dynamical model of error propagation in resilient Boolean Networks.

\begin{figure}[hbt!]
\includegraphics*{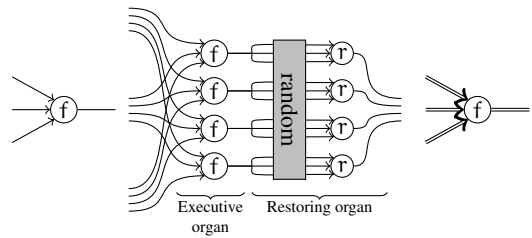}
\caption{Redundancy construction method. Left: The original function in the
network; Middle: Equivalent redundant function, composed of the executive and
restoring organs, with an edge bundle of size four. The grey rectangle
corresponds to a random rewiring of the edges; Right: The resulting
``pseudo-function'', with incoming and outgoing edge
bundles.\label{fig:restoration}}
\end{figure}

The mechanism proposed in~\cite{von_neumann_probabilistic_1956} consists of
locally replicating a given function, such that the replicated input and output
edges will form \emph{bundles} which will all carry the same information, in the
absence of noise (see Fig~\ref{fig:restoration}).
The edges of the output bundle are then randomized and fed into appropriate
restoration functions which will independently query the majority state carried
in them.  The number of edges in the bundle can be arbitrarily large, but the
number of inputs per node must remain fixed. The output bundle of the last
functions will then propagate the information to the rest of the network, which
is also modified in the same manner. The first stage was dubbed
in~\cite{von_neumann_probabilistic_1956} the ``executive organ``, and the second
stage, the ``restoring organ``.  We note that while this method outlines a
specific construction, it has a general nature, since it incorporates the two
the most necessary features to be resilient against noise: replication and
restoration of majority values. It does so piecewise for all functions in a
given network, depicting an alternative version with an optimal level of
redundancy.  This robust version will then function exactly like the original
network, if each executive and restoring organ is thought of as an individual
function (which we will call a \emph{pseudo-function}). However, in the presence
of noise, the fraction $b$ of outputs having value $1$ (and conversely $1-b$
with value $0$) that enter and leave such a pseudo-function is no longer a
Boolean value, but instead are \emph{real} values in the range $[0,1]$, which
will be continuously distributed in the limit of large number of edge in the
bundles. In the following, we assume that the number of edges in the bundle is
large enough, so that the fluctuations of the values of $b_i$ can be
neglected. In this case, these pseudo-functions will be generalizations of the
original Boolean function (plus restoration) in the real domain, which will
regulate how noise is propagated in the network. The general form of those
pseudo-functions is
\begin{equation}\label{equation}
  g (\{b_i\}) = r\left(\sum_{j=0}^{2^k-1}\delta_{f(j),1}\prod^{k-1}_{i=0}(\delta_{j_i,1}b_i+\delta_{j_i,0}(1-b_i))\right),
\end{equation}
where $\delta_{ij}$ is the Kronecker delta, $b_i$ is the input bundle $i$, $f$
is the function of the executive organ, the variable $j$ represents a given
Boolean input combination, $j_i$ the $i$-th bit of the input combination $j$,
and $r(b)$ performs the restoring function (which we will describe in detail
below).

We are interested in analysing how the above mechanism can hinder error
propagation on the network. The effect of noise can be measured in a variety of
ways, but here we are interested in the ability of the system in remembering its
past states, which we will label as non-ergodicity. More precisely, we can
define an order parameter, the long-term hamming distance,
\begin{equation}\label{eq:order_parameter}
  h = \lim_{T\to\infty} \frac{1}{T}\sum_{t=0}^T\left<\left|b_i(t|\sigma^a_i)-b_i(t|\sigma^b_i)\right|\right>,
\end{equation}
where $b_i(t|\sigma)$ is the value of $b$ for pseudo-node $i$ at time $t$, with
a starting state $b_i(0|\sigma)=\sigma$. The average $\left<\dots\right>$ is
taken over the whole network, and several independent realizations of the
dynamics, with different initial states $\{\sigma^a_i\}$ and $\{\sigma^b_i\}$.
If the system shows ergodic dynamics (as previously discussed), the value of $h$
should converge to zero, corresponding to only one possible fixed point in the
values of $b_i$. Otherwise, it should decrease with the noise strength $p$, as
the effects of noise brings the system closer to the ergodic phase~\footnote{We
note that those distinct phases are not in general related to the frozen or
``chaotic'' phases of random Boolean networks (RBN). In the presence of noise,
the concept of frozen or chaotic dynamics is not applicable. Furthermore,
without noise, a RBN either in the frozen or the chaotic phase can independently
be ergodic or not, according to the definition of ergodicity used in this
work.}.

We will consider separately the case of networks with functions $k=2$, and later
the case $k>2$. The case $k=1$ will not be analysed since it does not allow for
the construction of a restoring organ.

\begin{figure}[hbt!]
  \centering
  \includegraphics*[width=0.49\columnwidth]{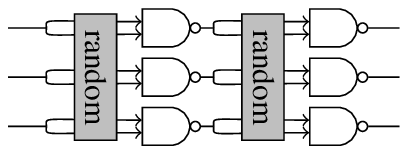}
  \includegraphics*[width=0.49\columnwidth]{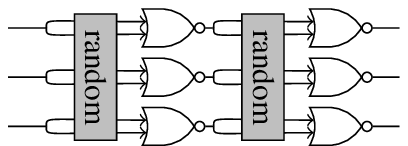}
\vspace{0.2cm}

\includegraphics*[width=0.48\columnwidth]{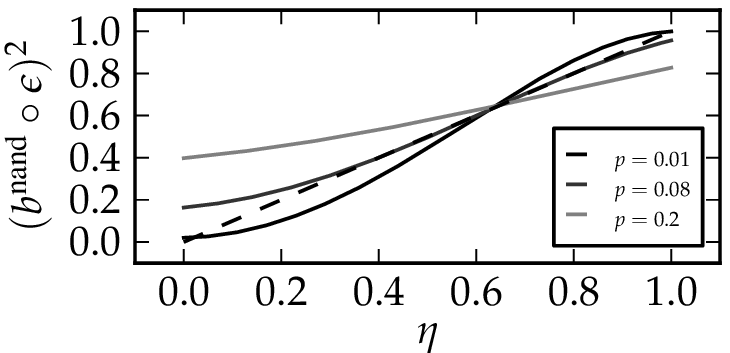}
\includegraphics*[width=0.48\columnwidth]{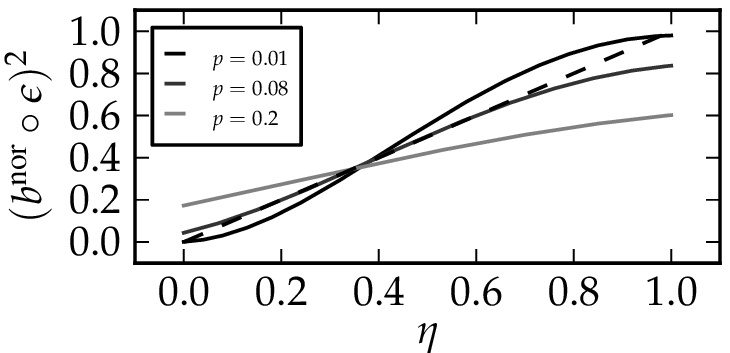}
\caption{Restoration organs with $k=2$, with NAND (left) and NOR (right)
functions. Below each are the restoration maps, for some values of $p$. The
dashed lines correspond to $b(\eta)=\eta$.\label{fig:restoration_k2}}
\end{figure}

\begin{figure}[htb!]
  \includegraphics*[width=0.45\columnwidth]{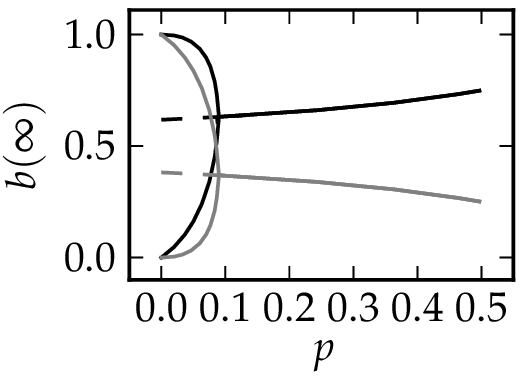}
  \includegraphics*[width=0.45\columnwidth]{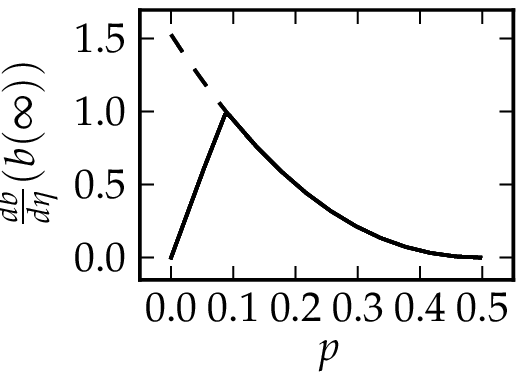}\\
  \hspace{-0.3cm}
  \includegraphics*[width=0.97\columnwidth]{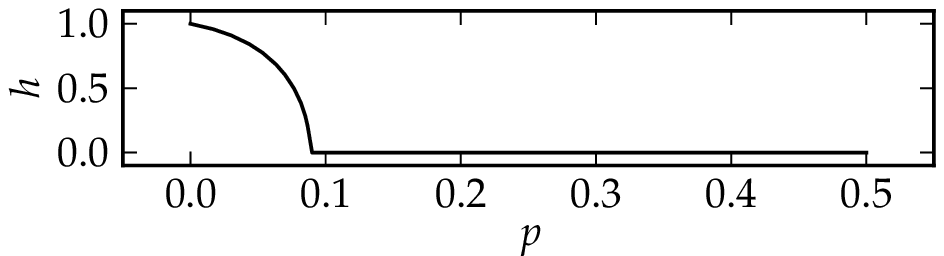}

  \caption{Fixed points, or period-2 points, $b(\infty)$ (top) and long-term
  hamming distance $h$ (bottom) for the NAND (black) and NOR (gray) maps, as a
  function of noise. Black curves correspond to NAND restoration and grey ones
  to NOR.\label{fig:nand-nor-transition}}
\end{figure}

\emph{$\mathit{k=2}$}--- The most crucial part in the procedure outlined above
is the selection of the function to be used in the restoring organ. The function
must be able to transform the values of the majority of the edges in the input
bundle, into a even greater majority in the output bundle. However, no $k=2$
function can act as a simple majority function. There are however some functions
which behave as a majority function for certain input combinations, but not for
others. These functions are the AND ($8$), NAND ($7$), OR ($14$) and NOR ($1$)
which react only to two simultaneous input changes, if the original inputs are
all $1$ or $0$, but react to any input flip if the original inputs are in the
opposite state. Therefore those functions would be able to correct either value
passing on the bundle, but not both. The solution proposed
in~\cite{von_neumann_probabilistic_1956} is to construct the restoring organ
with \emph{two} modules connected in sequence, both with either NAND or NOR
functions~\footnote{It is equivalent to use the AND or OR functions. We restrict
the analysis to the NAND and NOR functions without loss of generality.}, as can
be seen in Fig.~\ref{fig:restoration_k2}.
The first tier will correct one of the values if it can and intrinsically flip
the majority value of the bundle, and the second tier will then have its chance
of correcting, now that the majority value has changed. We can verify the actual
response of this scheme to noise, by defining two maps. First, the actual noise
on the bundle,
\begin{equation}\label{eq:noise}
  \epsilon(\eta) = (1-2p)\eta + p,
\end{equation}
where $\epsilon$ is the fraction of the edges in the bundle with a given value,
given the original fraction of same value $\eta$. Second, the response of the
NOR and NAND functions,
\begin{equation}\label{eq:or_map}
  b^{\text{nor}}(\eta) = (1-\eta)^2, \;\;  b^{\text{nand}}(\eta) = 1-\eta^2,
\end{equation}
where $b$ is the fraction of edges in the output bundle with value $1$, and
$\eta$ the fraction of inputs with the same value. The full restoration map is
then given by
\begin{equation}\label{eq:nand_restoration}
  b^r(\eta) = (b^\text{nor/nand} \circ \epsilon)^2(\eta),
\end{equation}
which is plotted in Fig.~\ref{fig:restoration_k2}, for some values of $p$. We
can see that the majority value on the bundle is preserved, even for non-zero
values of $p$. However, the question remains if this restoration will be enough
to maintain trajectories of a network from diverging. For that, we need to
couple the restoration map above with the functions present on the network and
iterate the system. However, there is one specific situation which represents
the limiting case of maximum resilience, namely when after each pseudo-function
there are infinitely many restoring organs in sequence. In this case, the
response to noise of the function in the executive organ can be neglected, and
the (infinitely long) restoring organ alone will determine the resilience of the
network. Conveniently, this can be done with successive iterations
Eq.~\ref{eq:nand_restoration}, which should eventually reach a fixed point,
corresponding to the roots of the equation $(b^{\text{nor/nand}} \circ
\epsilon)^2(\eta) = \eta$. This is a fourth order polynomial in $\eta$, and the
roots can be obtained analytically. The system exhibits a typical pitchfork
bifurcation, with three distinct fixed points (only two of each are stable), up
to a critical value of noise $p_c=(3 -\sqrt{7})/4\approx 0.0886$, above which
only one fixed point exists. This corresponds to a dynamical phase transition,
where below this critical point the values of the outputs will oscillate between
values close to $0$ and $1$, and thus memory of the past states will always be
preserved. Above the critical point, the dynamics is ergodic, independent of any
starting state. In order to characterize this transition more precisely, we can
write the expression for the previously defined order parameter in
Eq.~\ref{eq:order_parameter} as
\begin{align}\label{eq:h_nand}
    h &= \lim_{t\to\infty} [b(t|1) - b(t|0)] \\
      &= \frac{[8(p-p_c)(p-p_c^*)]^{\frac{1}{2}}}{(2p-1)^2}
    \qquad [p\leq p_c]
\end{align}
where $b(t|\sigma)$ is the value of $b(t)$ with the starting point $b(0) =
\sigma$, and $p^*_c=(3 + \sqrt{7})/4$. The values of the order parameter are
plotted in Fig~\ref{fig:nand-nor-transition}. From Eq.~\ref{eq:h_nand} we also
see easily that the critical exponent is $1/2$ (mean-field universality
class).
The existence of this critical value of noise points to a direct upper bound on
the reliability attainable by $k=2$ networks, since it represents the maximum
limit of error correction of the restoring organ. Additionally, this critical
value of noise corresponds exactly to the upped bound found rigorously by Evans
and Pippinger~\cite{evans_maximum_1998} for reliable computation of Boolean
formulas composed of noisy NAND gates.

\begin{figure}[htb!]
  \includegraphics*[width=0.45\columnwidth]{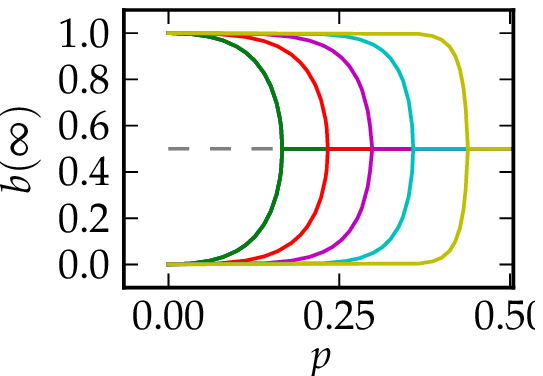}
  \includegraphics*[width=0.45\columnwidth]{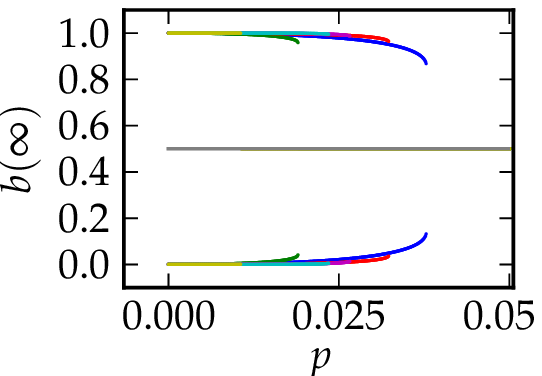}\\
  \includegraphics*[width=0.45\columnwidth]{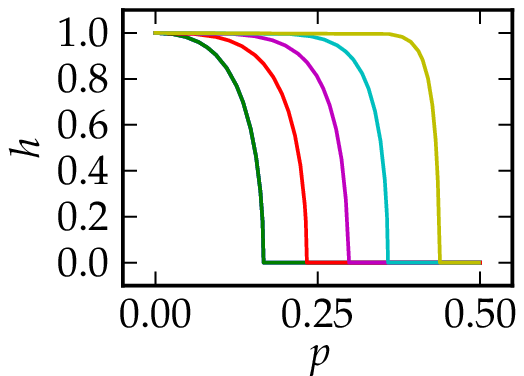}
  \includegraphics*[width=0.45\columnwidth]{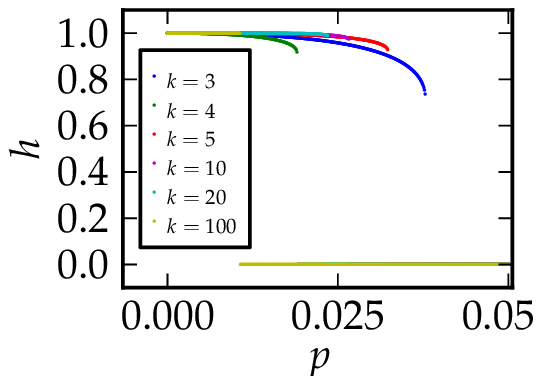}

  \caption{Fixed points and long-term hamming distance, for the majority
  restoring organ (left), and the maximum sensitivity executive organ with the
  majority restoring organ (right). \label{fig:majority_transition}}
\end{figure}

\begin{figure}[htb!]
  \includegraphics*[width=.45\columnwidth]{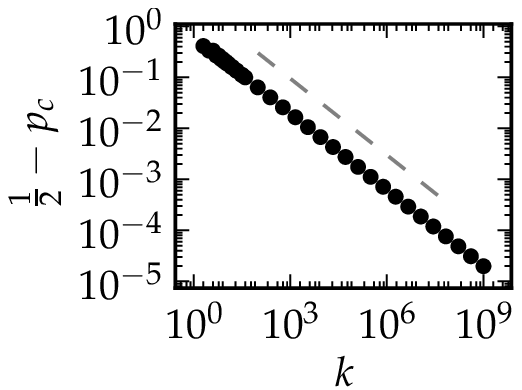}
  \includegraphics*[width=.45\columnwidth]{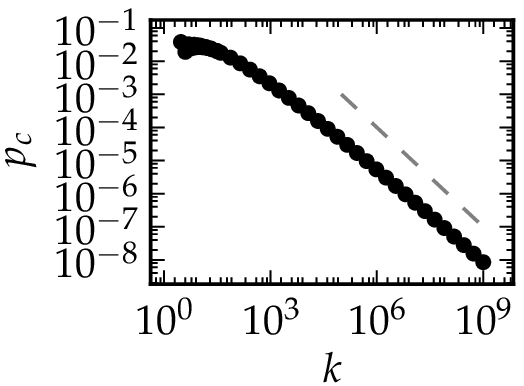}

  \caption{Critical noise value $p_c$ as a function of $k$, for the majority
  restoring organ, and $k=2$ NAND/NOR restoration (left), and the maximum
  sensitivity executive organ with the majority restoring organ (right). The
  dashed lines correspond to $1/\sqrt{k}$ (left) and $1/k$
  (right).\label{fig:majority_critical_p}}
\end{figure}

\emph{$\mathit{k>2}$}--- If the functions have $k>2$, the choice of the
restoring organ becomes more obvious, and the most natural choice is the
majority function, which returns simply the majority value of its inputs. Since
it will work equally well if the value on the bundle is either $0$ or $1$, the
majority function is capable of performing restoration with only one tier of
functions, without accumulating noise in an intermediate step, which provides it
with superior characteristics. The restoration map of the majority function is
given by
\begin{multline}\label{eq:majority}
  b^m(\eta) = 1-\sum_{i=0}^{\ceil{k/2}-1} {k\choose i} \eta^i(1-\eta)^{k-i} +\\
  \delta_{k/2,\floor{k/2}}\frac{1}{2}{k\choose k/2} \eta^{k/2}(1-\eta)^{k/2}.
\end{multline}
The last term is added only for functions with even $k$, which have an
indeterminate majority state. In this case, it is assumed that half the
restoring functions output $1$ and the other half $0$. It is also clear that
majority functions with even $k>2$ will perform just as well as a $k-1$ odd
function, and therefore the extra input is, for this purpose, wasted. We can
analyse the quality of this restoration by iterating Eqs.~\ref{eq:majority} and
~\ref{eq:noise} in sequence, like it was done for $k=2$. In the absence of
noise, this will lead to one of two fixed points, depending on the starting
condition. It can be seen in Fig.~\ref{fig:majority_transition} that those fixed
points also merge into one at a critical value of noise, and the associate order
parameter $h$ also indicates a second order transition, with the same critical
exponent, but different critical noise values. As expected, the value of $h$ is
larger for larger $k$, for the same value $p$, and the critical noise is also
larger. The critical values $p_c$ match exactly the upper bounds for Boolean
formulas using noisy majority functions of $k$ inputs found by Evans and
Schulman~\cite{evans_maximum_2003}, given by,
\begin{equation}
  p_c = \frac{1}{2} - \frac{2^{k-2}}{k{k-1 \choose \frac{k-1}{2}}},
\end{equation}
for odd $k$. This scales with $(1/2-p_c) \sim 1/\sqrt{k}$ (see
Fig.~\ref{fig:majority_critical_p}), leading to a strictly resilient situation
when $k\to\infty$.

One can test the effectiveness of the majority functions, by considering other
functions as the executive organ. Instead of systematically analysing all
$2^{2^k}$ functions, we can consider functions which are pathological in their
stability to noise. Here we will consider a function with maximum sensitivity,
which output $0$ or $1$ if \emph{all} inputs are $1$ or $0$, respectively, but
otherwise the function is uniformly distributed, and the outputs will be $0$ or
$1$ with equal probability for each input combination. The corresponding map can
be written as
\begin{equation}
  b^{\text{min}}(\eta) = \frac{1}{2}(1 + (1-\eta)^k - \eta^k)).
\end{equation}
Its properties can be seen in Fig.~\ref{fig:majority_transition}. We see that
indeed it becomes progressively difficult to stabilize for larger $k$, and the
critical point now scales as $p_c\sim 1/k$, and the transition becomes
first-order. On the other hand, the mere existence of the critical point
confirms some level of resilience, which is not present in the system without
redundancy, where the critical point is always $p_c=0$.

\emph{Conclusion}--- We have shown that sparse networks, while they cannot be
arbitrarily resilient, they can have stable dynamics in the presence of noise,
if redundancy is correctly introduced. The stability is marked by second or
first-order transitions, from non-ergodic to ergodic behaviour. We obtain
upper-bounds on the error resilience attainable by redundant networks with a
given $k$. This is in stark contrast to what is observed in Random Boolean
Networks~\cite{peixoto_noise_2009}, which never exhibit memory of its past
states when noise is introduced, either in its frozen or ``chaotic'' phases.

We have shown that the stabilization through redundancy is successful even with
the most pathologically sensitive functions, such as the function with maximum
sensitivity discussed. We note also that redundancy provides additional
benefits, such as robustness against damage and mutations, as was shown
in~\cite{gershenson_role_2006}.  We stipulate that due to these robust features,
redundancy must be present in some extent in real gene regulatory networks; if
not in the entire network, at least in its more dynamically relevant modules.
On the other hand, arbitrary redundancy close to the optimal bound is very
unlikely due to its high putative cost to the organism, which would favor
instead a genetic circuit composed of fewer elements, with only enough
resilience sufficient for survival. It remains to be seen to what extent is
redundancy desirable, and how it may be connected with other topological and
functional restrictions of gene regulation.

\emph{Acknowledgement}--- I thank Tamara Mihaljev and Barbara Drossel for
carefully reviewing the manuscript, and for suggestions. This work has been
supported by the DFG, under contract number Dr300/5-1.

\bibliography{bib}

\end{document}